# The Knowledge Gap in a High-Choice Media Environment: Experimental Evidence from Online Search

Ulloa, Roberto[1]; Leonard Tiedemann[1,2]; Selb, Peter[2]; Kacperski, Celina[1,3]*

[1] Cluster of Excellence "The Politics of Inequality", University of Konstanz, Konstanz, Germany
[2] Department of Politics and Public Administration, University of Konstanz, Konstanz, Germany
[3] Cluster of Excellence "Collective Behaviour", University of Konstanz, Konstanz, Germany

*Corresponding author: celina.kacperski@uni-konstanz.de

## Statements and Declarations

**Acknowledgments**
We thank the Seek2Judge team for their support.

**Funding**
Deutsche Forschungsgemeinschaft (DFG – German Research Foundation) under Germany's Excellence Strategy – EXC-2035/1 – 390681379

**Competing interests**
The authors declare that they have no competing interests.

**Data Availability Statement**
Data and analysis scripts are available at:
https://osf.io/pv8ey/files/osfstorage?view_only=f86afcdc082b4814beed8130a9f33d5c

**AI usage statement**
This manuscript made use of *ChatGPT (GPT-4 and GPT-5)* to assist with language polishing and consistency, and writing clarity improvements. The authors reviewed and edited all outputs, and they take full responsibility for the final text.

## Abstract

Persistent inequalities in political knowledge are a central concern in political communication. We organize the mechanisms underlying the knowledge-gap literature by distinguishing between individual preconditions, structural features of the information environment, and topic characteristics. Within this framework, we note that self-directed information seeking, a prototypical form of intentional exposure, has received little attention despite its importance in navigating today's complex information environment. We conducted a field experiment in Germany combining randomized encouragements and passive browser tracking to examine how individuals with varying education levels acquire policy-specific knowledge through online search. Participants were randomly assigned to one of three conditions (verbal encouragement, financial encouragement, or control) to seek information on three salient policy topics differing in divisiveness and complexity (child support, energy transition, and cannabis legalization). We estimate both intention-to-treat (ITT) and local average treatment effects (LATE) of information seeking on post-search knowledge outcomes, with a focus on education and civic knowledge as moderators. While the interventions equalized information-seeking behavior, the results provide some support for the knowledge gap hypothesis: knowledge gains were concentrated among participants with higher education or baseline civic knowledge, who, according to our post-hoc exploratory analyses, appeared more effective at navigating search results. These findings indicate that a narrowing of knowledge inequalities goes beyond motivation: it calls for both individual-level interventions to strengthen citizens' skills and structural-level adaptations to foster more equitable learning environments.

## Introduction

Political knowledge is a cornerstone of democratic participation, influencing aspects ranging from citizens' opinions and engagement to their voting decisions (Carpini & Keeter, 1996; Prior, 2007; Tichenor et al., 1970; Van Aelst et al., 2017). Yet decades of research show that access to political information does not translate into equal learning outcomes (Lind & Boomgaarden, 2019). Instead, better-educated individuals often learn more, and more quickly, from the same media content, a phenomenon known as the knowledge gap hypothesis (Tichenor et al., 1970). As media systems evolve, so too do the mechanisms that underlie these disparities.

The knowledge gap literature has identified many mechanisms that lead to learning inequalities. In Figure 1, we conceptualize these mechanisms as, either, individual-level characteristics (left block) associated with educational levels, such as motivation, cognitive ability, and prior knowledge (Kwak, 1999; Price & Zaller, 1993; Tichenor et al., 1970; Viswanath et al., 1993; Viswanath & Finnegan, 1996), or structural-level conditions (right block), including the type of media used, the publicity and duration of coverage, and unequal access to information (Gaziano, 1983; Kwak, 1999; Tichenor et al., 1970). More recent work extends these structural conditions to digital contexts (for a review, see Lind & Boomgaarden, 2019), where algorithmic mediation and high-choice environments reshape how individuals encounter and interact with information (Prior, 2007; Van Aelst et al., 2017). These features of digital media systems affect pathways to political learning that have become central to recent research. In particular, researchers have been recently conceptualizing the connections between incidental, intentional, and selective exposure (Kwak et al., 2022; Matthes et al., 2020; Schäfer, 2023; Thorson, 2020). We build on this perspective by noting that individuals ultimately decide whether to engage with the content they are exposed to, meaning that mechanisms such as selective exposure may operate regardless of whether the initial encounter is intentional or incidental.

**Figure 1. Conceptual framework of political learning outcomes.**

**Individual**
Communication skills
Prior knowledge
Social connections
Motivation / interest (*)
*Online search skills (**)*

**Topic features**
Temporality
Specificity
Complexity
Divisiveness

**Structural**
Media type
Media access
Publicity / time
*High-choice environment (**)*
*Algorithmic mediation (**)*

Incidental / intentional exposure

Selection mechanisms (e.g., selective exposure)

Learning outcomes

Notes: Learning outcomes result from the interaction between inherent characteristics of the topic (top block), individual- and structural-level mechanisms (left and right blocks, respectively) that shape how political information is encountered and processed. These mechanisms influence whether individuals are exposed to information intentionally or incidentally, and whether they subsequently engage with and learn from it. (*) Motivation is directly targeted by our intervention. Elements in italics marked with (**) represent relevant features of digital media environments.

[**Alt Text:** Conceptual diagram showing how topic features, individual-level mechanisms, and structural conditions shape incidental or intentional exposure, selection of information, and political learning outcomes.

**Long description:** A flow diagram linking political learning to three upstream inputs and two intermediate stages. At the top are three blocks. The first block lists topic features: temporality, specificity, complexity, and divisiveness. The second block lists individual-level mechanisms: communication skills, prior knowledge, social connections, motivation and interest, and online search skills. The third block lists structural-level conditions: media type, media access, publicity and time, high-choice environments, and algorithmic mediation. Arrows from all three blocks point to a stage describing how political information is encountered, distinguishing incidental from intentional exposure. A second arrow leads to a selection stage describing whether individuals engage with encountered content, for example through selective exposure. A final arrow leads to political learning outcomes. The diagram highlights that the intervention targets motivation within the individual-level mechanisms block, and that online search skills, high-choice environments, and algorithmic mediation are specific to digital media contexts.]

In this conceptual framework, learning outcomes are best understood as the result of interactions between individual predispositions, structural conditions, and inherent characteristics of the topic (center block of Figure 1). The interplay is particularly salient in the case of online search, a prototypical form of intentional exposure in which individuals deliberately seek (political) information to satisfy specific informational demands. While incidental exposure has been studied as a potential knowledge equalizer (Baum, 2002; Kobayashi & Inamasu, 2015; Li & Cho, 2023; Weeks et al., 2022), actively seeking information remains the dominant pathway that reflects the role of individual agency in navigating political information online.

Therefore, our study focuses on online search as a form of intentional exposure, a behavior that is central to navigating high-choice, algorithmically curated media environments. We investigate whether simple, low-cost interventions can shape search behavior and affect knowledge inequalities. Specifically, we examine whether prompting individuals to search for information affects how they seek information and what they learn from it, and whether these effects vary by educational background.

## The Knowledge Gap Hypothesis

The knowledge gap hypothesis states that as new information is disseminated via media, individuals with higher levels of education assimilate this knowledge more rapidly than those with less education (Tichenor et al., 1970). Tichenor et al. (1970) argue that widening knowledge gaps arise from individual predispositions (skills, prior knowledge, social ties, and motivationally driven selective exposure) interacting with structural characteristics of mass media.

Since then, researchers have identified other features of the media environment that determine how information is distributed, framed, and accessed, such as media types (e.g., online, printed, television), the availability and cost of access, and the amount of publicity surrounding political issues (Lind & Boomgaarden, 2019). Digital environments have brought new structural considerations: a high-choice media environment that allows users to select from an abundance of political and non-political content, increasing the risk of fragmentation (Prior, 2005; Van Aelst et al., 2017), and algorithmic curation that shapes the visibility and salience of political information in ways that vary across users (Thorson, 2020). At the same time, a cross-cutting factor of the knowledge gap is the knowledge type that is being evaluated (center block of Figure 1): gaps emerge most consistently for time-sensitive, policy-specific, non-contested, and complex issues (Barabas et al., 2014; Kleinnijenhuis, 1991; Lind & Boomgaarden, 2019).

**Navigating Political Information in Digital Environments**

While the knowledge gap hypothesis originally focused on differential learning from exposure to analog mass media (Tichenor et al., 1970; Viswanath & Finnegan, 1996), digital environments have transformed how individuals are exposed to (political) information. Due to the explosion of available information in the last decades (Prior, 2005, 2007; Van Aelst et al., 2017), researchers have become increasingly interested in incidental exposure, i.e., encounters without intentionally searching for them (Nanz & Matthes, 2022), and selective exposure, the selection of information matching beliefs (Knobloch-Westerwick, 2014). These mechanisms often operate in sequence or combination (Kwak et al., 2022; Matthes et al., 2020; Schäfer, 2023; Thorson, 2020).

Unlike incidental exposure, online search is a prototypical case of intentional behaviour that reflects the user's initiative to fulfill a need or goal (Case & Given, 2016). It is intrinsically associated with those who are motivated to learn, who know how to search effectively, and who ultimately benefit from digital resources (Hargittai & Hinnant, 2008; Scheerder et al., 2020; van Deursen & van Dijk, 2014).

That said, search is not merely an individual act; it is also shaped by structural constraints. Search engines rank, filter, and label information in ways that affect what users see, regardless of their *intent* (Noble, 2018). Still, the individual ultimately preserves agency in terms of the search platform selection, prompt formulation, and the final selection of the sources they engage with;

their communication skills, prior knowledge, and intrinsic interest (left block of Figure 1) can affect those decisions and, ultimately, the learning outcomes. Thus, our study centers on self-directed search as a behavior that can either mitigate or exacerbate learning inequalities.

**Knowledge Inequalities in Digital Environments**

Research has indeed shown that individuals with higher levels of education use the internet more frequently for seeking (political) information (Dijk, 2020; Hargittai & Hinnant, 2008; Scheerder et al., 2020; van Deursen & van Dijk, 2014), including for political information (Scheerder et al., 2020; van Deursen & van Dijk, 2014), are better able to navigate and benefit from online resources (Hargittai & Hinnant, 2008), and devote more time to informational rather than entertainment content and to consulting background material (Scheerder et al., 2020; van Deursen & van Dijk, 2014), which contributes to widening education-based knowledge gaps as shown in longitudinal (Gibson & McAllister, 2015; Wei & Hindman, 2011), experimental (Yang & Grabe, 2011, 2014), and social media studies (Woo-Yoo & Gil-de-Zúñiga, 2014), with reviews concluding that online media amplify knowledge gaps more than print (Lind & Boomgaarden, 2019).

However, contrary to the knowledge gap hypothesis, some research also suggests that online and social media can narrow knowledge inequalities (Cacciatore et al., 2014; Norris, 2003), with correlational evidence showing smaller gaps among frequent internet users and digital news consumers (Anduiza et al., 2012; Lee & Yang, 2014), largely attributed to incidental exposure through soft news and entertainment contexts (Baum, 2002; Kobayashi & Inamasu, 2015; Morris & Morris, 2017), social media use (Li & Cho, 2023), and low-interest learning from incidental exposure (Weeks et al., 2022), supported indirectly by studies linking information access to political knowledge in restricted media systems and cases where alternative online media reduced education-based gaps (Chen & Yang, 2019; Goh, 2015; Miner, 2015; Placek, 2020).

Additionally, some studies, such as one on cross-national evidence from 1995–2015 in most democracies (Haugsgjerd et al., 2021), claim no change in education-based knowledge gaps, and suggest stable knowledge inequalities due to offsetting mechanisms: for example, incidental exposure can yield equal gains across media preference groups unless individuals self-select their preferred content (Leeper, 2020) and ceiling effects or periods of intense information flow such as campaigns may further limit gap growth (De Silva-Schmidt et al., 2022; Ettema & Kline, 1977; Hansen & Pedersen, 2014; Kwak, 1999).

Given the complexity of the current media landscape, experimental designs have been used to disentangle causal mechanisms of the knowledge gap, for instance, the role of type of media (Grabe et al., 2009; Yang & Grabe, 2011), knowledge domains (Yang & Grabe, 2014), emotionality (Bas & Grabe, 2015) and incidental exposure (Leeper, 2020), with the latter focusing on media preferences rather than education as the key individual-level characteristic.

Building on these experimental contributions, it is also important to distinguish how our design relates to and extends prior methods that address different aspects of political learning. Prior & Lupia, (2008) experimentally showed that conventional survey contexts confound respondents' quick recall with motivation and opportunity to engage with political information, finding that providing incentives and extra time increases correct responses by reducing motivational barriers to retrieval and effortful thought. While we similarly ensure motivation and preparation time across conditions, we do not manipulate test conditions per se, instead we leverage a realistic online search period followed by timed assessment to examine what participants have learned through self-directed search, not how survey features affect performance. Other work has randomized assigned exposure to particular outlets or content, such as studies encouraging participants to follow specific media sources and then testing associated knowledge gains (Altay et al., 2025; Broockman & Kalla, 2025; Guess et al., 2021); these designs are useful for isolating the effects of forced content exposure, but they neither capture the processes of active information seeking in high-choice environments nor their direct role in the knowledge gap as we propose here. Also, laboratory designs that emphasize media choice and self-selection, such as those grounded in the selective exposure paradigms (Leeper, 2020), focus on how allowing participants to choose content shapes outcomes, often for opinion or attitude measures, and are framed around media preference as a driver of heterogeneity rather than the mechanism of motivated search itself. In contrast, our study integrates actual browsing behavior over an extended period with controlled motivational conditions and subsequent knowledge outcomes, thereby linking the causes of search behavior to its consequences for learning in ways that neither survey-based incentive experiments nor forced exposure and choice designs fully address.

Finally, a strand of work uses web browsing and tracking data to observe actual online information environments and user behavior: recent advances in browsing-level trackers (e.g., Adam et al., 2024) allow researchers to directly capture the content people encounter in situ, offering richer behavioral data than surveys or self-reports and helping examine how real browsing patterns relate to exposure and knowledge. For example, browsing data has previously been used to assess knowledge differentials as a function of relative preference for entertainment over news (Kobayashi & Inamasu, 2015). While such web-tracking studies provide important observational insight into differential online behavior, they typically do not experimentally manipulate search motivations as we do; thus, our design uniquely combines controlled motivational intervention conditions with detailed behavioral traces and knowledge outcomes, linking search behaviors to its learning consequences in contemporary online environments beyond purely observational evidence.

## Research Agenda

This study advances knowledge‑gap research by testing how individuals with different education levels acquire policy‑specific knowledge through self-directed (intentional) online search in a contemporary high-choice media environment. We ask whether encouragements and actual search widen or narrow knowledge gaps (RQ1, RQ2), and explore variation in search compliance and effort (RQ3, RQ4). From the individual mechanism of our conceptual framework (left block of Figure 1), the encouragement targets individuals' motivation to learn about the topic via search. Using randomization (of encouragements) and behavioral traces of search, we estimate both the intention-to-treat (ITT) effects of assignment and the local average treatment effect (LATE) of actually searching. We operationalize education levels as either formal educational attainment (low vs. high) and civic knowledge, i.e., pre‑existing general political knowledge, measured with a set of party–politician allocation items validated for the German context (Moosdorf et al., 2020). While correlated (Kendall's $\tau = .15$, $p < .001$), education reflects structural access to general resources, skills, and motivation, whereas civic knowledge captures these capacities as applied within the political context, independent of structural background. Our design includes: (i) individual differences via education and baseline civic knowledge as well as motivation via encouragements; (ii) structural realism by allowing participants to search in the wild (we do not prescribe engine, query, or source, thereby embedding real-world algorithmic curation), and (iii) three salient policy-topics, varying on their divisiveness and complexity.

On the confirmatory research side, our first question focuses on the intention-to-treat (ITT) effect: ***RQ1.*** *Does encouragement to inform oneself online about a policy affect policy knowledge differently across individuals with varying educational levels?* ***H1a.*** *Participants who received a verbal or monetary encouragement treatment will show higher scores on the knowledge test compared to participants in the control group.* ***H1b.*** *Knowledge test scores will differ between those with higher educational levels compared to those with lower educational levels. Individuals with higher education levels will have higher test scores.* ***H1c.*** *The difference in knowledge test scores for those who received an encouragement treatment (compared to those who did not) will differ for individuals with varying education levels, i.e., the ITT effect will be moderated by education.*

We then turn to a local average treatment effect (LATE) analysis, isolating the impact of actually conducting a search: ***RQ2.*** *Does online information search affect individuals' policy knowledge differently across individuals with varying educational levels, i.e., does it affect the knowledge gap? (LATE effect).* ***H2a.*** *Participants who informed themselves about a policy online will show higher knowledge test scores compared to participants who did not conduct an information search.* ***H2b.*** *The difference in knowledge test scores for those who informed themselves online about the policy (compared to those who did not) will differ for individuals with varying education levels, i.e., education will moderate LATE.*

Both H1c and H2b are specified without a directional forecast. Although the knowledge gap hypothesis leads us to expect that more educated individuals will gain more, thus widening disparities, it is also plausible that lower-educated participants, starting from a lower baseline, could achieve relatively larger gains when prompted to search, as it might alter mechanisms, such as motivation, that underlie the knowledge gap hypothesis.

Beyond these confirmatory tests, we explore how encouragement and education shape search behavior: ***RQ3.*** *Does encouragement influence the likelihood of conducting an online search across educational levels? (i.e., compliance effect).* ***RQ4.*** *Do search-effort metrics (e.g., number of pages visited, time spent per search) vary by educational level?*

## Methods

The study was conducted in Germany, a multi-party parliamentary democracy with a mixed media environment in which strong public service broadcasters operate alongside a diverse and increasingly digital news ecosystem (Behre et al., 2025). The experiment occurred in 2023, a non-election period. In this context, political debates are not driven by a single central campaign but by ongoing parliamentary processes and varying media attention cycles. This environment allows for the observation of information-seeking behavior outside of intense mobilization typical of election campaigns. To capture these dynamics across different levels of public debate, we selected policy issues that varied in their complexity, divisiveness and salience.

### Selection of policy topics and knowledge test items

To select the policies, we first pre-selected thirteen policies that were being discussed in the German parliament[1] and the news. In February 2023, we invited German individuals on the survey platform Prolific to participate (~€1/5min). A total of 83 individuals (42 women, 2 diverse) with a mean age of 37.31 years (SD = 13.64) rated the policies across four dimensions: saliency, divisiveness, complexity, and interest. We selected three policies according to the following criteria: being some of the most salient German-level policies (i.e., surveillance policy-specific topics, see Barabas et al., 2014) that would diverge in terms of divisiveness (for results regarding contested vs non-contested issues, see Lind & Boomgaarden, 2019) and complexity (Kleinnijenhuis, 1991), while being considered interesting (Figure 2). We selected (1) basic child support (Kindergrundsicherung: low divisiveness, medium to low complexity), (2) renewable energy transition (Förderung erneuerbarer Energien: high divisiveness, more complex), and (3) cannabis legalization (Cannabislegalisierung: high divisiveness, low complexity). For each of the three selected policies, we developed eleven items assessing knowledge related to the content of the policies. The policy item contents were revised by the authors and research assistants, who discussed the questions and verified the answers based on multiple trustworthy sources. Then, in May 2023, we again invited German individuals from Prolific to participate. A total of 89 (46 women, 2 diverse, mean age of 39.28 years, SD = 12.88 ) completed the items. We selected five items based on a Rasch-model analysis, prioritizing those with higher difficulty parameters to reduce the risk of ceiling effects. All items with their estimated difficulty (and selection) are included in Appendix B.

[1] We included two topics that were already ratified (Nuclear Exit Policy and Citizens' Income Benefit) and one discussed at the the European Union level (EU Fit for 55).

**Figure 2. Policy topic selection.**

| | Saliency | Divisiveness | Complexity | Interest |
|---|---|---|---|---|
| **Cannabis Legalization Act** | 74 | 64 | 28 | 64 |
| *Nuclear Exit Policy (Energiewende Phase-Out)* | 70 | 68 | 30 | 66 |
| *Citizen's Benefit (Bürgergeld, BG)* | 60 | 59 | 44 | 62 |
| **Renewable Energy Sources Act (Erneuerbare-Energien-Gesetz, EEG 2023)** | 60 | 58 | 54 | 64 |
| Basic Tax Reform (Grundsteuerreform) | 51 | 36 | 62 | 32 |
| **Basic Child Allowance (Kindergrundsicherung, KG)** | 46 | 26 | 34 | 52 |
| Long-Term Care Support & Relief Act (Pflegeentlastungsgesetz) | 30 | 27 | 51 | 53 |
| Digital Healthcare & Health Data Use Acts (DigiG & GDNG) | 28 | 22 | 44 | 56 |
| Hospital Care Improvement Act (Krankenhauspflegeentlastungsgesetz) | 24 | 22 | 49 | 52 |
| Private Pension Reform (Aktienrente) | 19 | 31 | 60 | 49 |
| *EU Fit for 55 Climate Package Implementation* | 18 | 61 | 54 | 54 |
| Public Services Protection Measures | 14 | 32 | 39 | 52 |
| Critical Raw Materials Strategy | 13 | 27 | 60 | 51 |

Note: Heatmap displaying median scores across four dimensions for 13 German policies. Policies are ranked by saliency (highest to lowest) and assessed on saliency (orange), divisiveness (purple), complexity (red), and interest (blue). Each cell shows the median score, with the color gradient representing the scaled score restricted to the min and max median in each dimension. On the X-axis, bolded policies represent the finally selected items, and policies in italics and gray were excluded from the final selection because they were already ratified or were related to the European Union.

**[Alt Text:** Heatmap of thirteen German policy topics rated on saliency, divisiveness, complexity, and interest; colors show scaled median ratings, and three highlighted topics are selected for the study: child support, renewable energy transition, and cannabis legalization.]

### Main survey design and data collection

In February and May 2023 members of a web tracking online commercial panel (Bilendi GmbH, which adheres to EU GDPR regulations) were invited to take part in a 20-minute baseline survey that queried their socio-demographics, political interest, and civic knowledge as a distinct moderator targeting political informational readiness (measured via validated items of general political knowledge in the German context, Moosdorf et al., 2020), among other baseline measures such as browsing preferences not used in this study. In June 2023, all individuals who completed the initial survey were re-contacted to participate in the experimental trial. The experiment was fielded in three topic-specific waves (child support, energy transition, cannabis legalization), each implemented as a separate two-part survey (pre-treatment questionnaire → ~20-hour search window → post-treatment questionnaire). Each wave was fielded in a different week, within a one-month window. Panelists could participate in any subset of these waves. In each wave, participants responded to a set of items regarding the policy, including attitudinal questions such as: “What is your attitude towards the introduction of basic child protection?” (7-point Likert scale ranging from (1) ‘should not be introduced’ to (7) ‘should be introduced’). After answering the initial set of questions, participants were randomly assigned to one of three experimental conditions. The control group received no additional instructions. The verbal encouragement group was prompted: “*Please inform yourself thoroughly in the next 20 hours on the internet about the current policy propositions regarding policy topic”*. The topic changed according to the wave. The monetary encouragement group received a similar prompt with an incentive: “*Please inform yourself thoroughly in the next 20 hours on the internet about the current policy propositions regarding policy topic. In the second part of the survey, we will* *conduct a knowledge test with five questions (for example: Example Question). If you answer at least 3 questions correctly, you will receive an additional xx mingle points.*" Mingle points are the panel company's mode of financially compensating the panelists and were equivalent to three euros. The example question was given as an illustration and did not appear in the final knowledge test. Participants were instructed to return 20 hours later for the postline survey. A reminder email was sent at that time. The 20-hour window provided a naturalistic opportunity for engagement (e.g., overnight reading) while minimizing attrition and confounding from external events (by keeping the field period short). Theoretically, this duration prevents underestimating knowledge by avoiding the surprise of an unexpected test within a survey (Prior & Lupia, 2008). We included modest financial incentives, alongside verbal encouragement, as part of an encouragement design commonly used in political communication research to promote task engagement, including recent media-effects experiments that incentivize exposure and verify learning through knowledge quizzes (Broockman & Kalla, 2025; Prior & Lupia, 2008).

After the treatment period, all groups then completed the postline survey, including the attitudinal questions regarding the policy, followed by the knowledge test. Participants were given 25 seconds to answer each of the 5 knowledge questions (Appendix B). The knowledge questions were only administered on the postline survey to avoid effects that could have led participants to search for those specific answers during the treatment period (civic knowledge captures baseline general political knowledge). The entire survey also included five attention test questions throughout (3 in baseline, 2 in postline). Survey items can be found in Appendix D.

**Participants**

From those German-speaking adults residing in Germany who completed the preliminary panel survey (N = 2,705), 1,216 participated in at least one wave of the experimental study. Since participation was allowed in each wave regardless of prior engagement, sample sizes varied by wave (Wave 1: N = 1066; Wave 2: N = 950; Wave 3: N = 954). Participants who entered the experiment were only slightly more politically interested and more civically knowledgeable than baseline-only respondents, and were marginally older and less rural; educational attainment and the remaining observed covariates did not differ statistically (Appendix A). Of those who entered the survey, 871 participants were (web) *tracked*, i.e., had at least one web visit during the experiment (Wave 1: N = 702; Wave 2: N = 637; Wave 3: N = 620). Demographic variables (education, age, gender, children, region, and income), political interest (5-point Likert scale), number of attention checks passed, and recruitment month (February or May) remained well balanced across treatment groups within each wave and for the participants with or without digital trace data; see Appendix A. The gender breakdown of the final sample included 568 women (46.7%) and 648 men (53.3%), overrepresenting men compared to national population figures (51.7% female). No participants identified as non-binary or other gender identities. Ages ranged from 18 to 80, with a median age of 49 years (mean = 48.1, SD = 12.3), older than the German adult population (median = 44.7 years). Educational attainment was categorized into two groups: low (i.e., from elementary school to having completed a middle school diploma; 47.8%), and high (completed a degree after middle school diploma; 52.2%). Compared to national statistics (Statistisches Bundesamt, 2024), the sample included a higher proportion of higher education individuals (52.2% vs. 39%). On average, participants correctly answered 6.76 (out of 11) civic knowledge questions (M=61.5%, SD = 25.8). Educational attainment and civic knowledge had a weak-to-moderate positive association (Kendall's $\tau = .15$, $z = 5.93$, $p < .001$).

**Processing of browsing data**

The software installed on the (desktop and mobile) devices of the panelists captured browser activity at the URL level, including timestamps and durations. Multiple *views* of the same URL could occur, for instance, if a browser tab is activated twice. These URL-level *views* were aggregated into URL *visits*, defined as sequences of the same URL accessed within a 30-minute window. During the 20-hour intervention periods, the 871 tracked participants generated ~452K visits (M = 439.8, Mdn = 257, SD = 520.65) corresponding to ~267K unique URLs. These URLs were manually annotated to determine whether they are related to the policy topics instructed in the intervention; we used the approach described in a recent study using the same dataset (Kacperski et al., 2025); see Appendix J. To evaluate the accuracy of the manual annotation procedure, we applied a machine learning classifier (BERT-large, F1-score > .97 across topics) to the entire URL set (for training details, see Schelb et al., 2024) . The classifier was tuned to maximize recall: a page was considered topic-related if any section of its content or URL suggested relevance. Manual inspection of the ~2,000 URLs identified as relevant by the classifier but not by manual annotations revealed that only six were missed by the manual annotation workflow and were added to the analysis. Across topics, we logged 1,886 policy-related visits, of which 930 were coded as searches (identified by the presence of a query term within the URL) and 956 as URL page visits, corresponding to 808 unique individual–search combinations and 860 unique individual–page visit combinations.

**Data analysis**

The main dependent variable of our study is the score on the policy knowledge test, i.e., the average correct answers. The analyzed variables are the experimental condition, treated as a categorical factor (control, verbal prompt, monetary incentive), information search (a binary, 1 indicating whether the participant visited a policy-related URL, 0 otherwise), educational attainment (low vs high), and civic knowledge. We estimate models separately by policy wave because the outcome measures are policy-specific and not directly comparable across topics.

To assess the intention-to-treat (ITT) effects (RQ1), we use linear regression models employing ordinary least squares (OLS) estimation (via the g*lm* function with a Gaussian identity link); we estimate the effect of the experimental condition moderated by educational attainment and by civic knowledge, i.e., fitting two models with interaction terms accordingly. Across all waves, power calculations indicate that we had >98.5% power to detect small effects (Cohen's f = 0.15). We also fit two analogous models using information search as the dependent variable (using glm with a logit link, RQ3).

While only the ITT effect is immediately identified from random assignment, under certain conditions (Sovey & Green, 2011), we can also estimate the local average treatment effect (LATE) of information search, i.e., the average effect of information search among those whose behavior was altered by the encouragement (Angrist & Imbens, 1995). The fulfillment of these conditions is discussed in Appendix C. Regarding the exclusion restriction, it is possible that the encouragements increase policy learning not only through purposive online search as reflected in our browser-tracking measures, but also through unobserved channels (e.g., interpersonal discussion or offline news). Accordingly, LATE estimates should be interpreted as the effect of encouragement-induced information acquisition insofar as it is captured in our observed search behavior.

For the estimation of the LATE, we use two-stage least squares (via the *ivreg* function, Fox et al., 2024), using the experimental condition as the instrument and information search as the endogenous variable. The analysis is restricted to the 871 individuals with web track data (see section participants); we verified that participants with and without digital trace data did not differ significantly on ITT outcome (Appendix E). We estimate two interaction models to test whether the effect of information search is moderated by educational attainment and civic knowledge, respectively. Both information search and its interaction with the moderator are treated as endogenous and instrumented using the random assignment and its interaction with the moderator, with the moderator included in both stages (Angrist & Pischke, 2009); instrument strength is assessed using first-stage F-statistics, reported separately for the main endogenous regressor and the interaction term. For the Cannabis Legalization interaction model, the instrument was weak, limiting confidence in the estimate of treatment effect heterogeneity, so the results should be interpreted with caution; when we discuss our results, we refer to the null effects obtained in the ITT/OLS regression, and not the null effects obtained in the instrumental variable regression.

For robustness, we also fit the regressions with controls for methodological variables (number of attention checks passed, participation in previous waves, and recruitment month) and other covariates (age, gender, children, region, income and political interest). As results remained consistent across specifications, we report the models without covariate adjustments; tables including controlled variables are provided in Appendix E to G. All analyses were carried out using R (R Core Team, 2024).

## Results

In the main effect models, the verbal and monetary encouragement have an effect on the knowledge score (H1a) for all waves except the verbal encouragement in the energy transition wave. We also find that both higher educational attainment and greater civic knowledge are positively associated with knowledge scores (H1b). We find partial support for the knowledge gap hypothesis (H1c). For educational attainment, we find statistically significant interactions with monetary encouragement for child support. For civic knowledge, we find significant interactions with verbal encouragement and with monetary encouragement for child support, as well as with monetary encouragement for energy transition. Table 1 presents the summary results.

**Table 1. Summary of intention-to-treat (ITT) regressions with and without interaction terms.**

| | Child Support (N=1066) | Energy Transition (N=950) | Cannabis Legalization (N=954) | Child Support (N=1066) | Energy Transition (N=950) | Cannabis Legalization (N=954) |
|---|---|---|---|---|---|---|
| ***Section A - Moderator: Educational attainment*** | | | | | | |
| ***Educ.*** | .06*** [.03, .09] | .08*** [.05, .10] | .06*** [.03, .09] | .02 [-.03, .06] | .09*** [.04, .13] | .03 [-.01, .08] |
| ***Verb. Enc.*** | .05** [.02, .08] | .00 [-.03, .04] | .06*** [.03, .10] | .02 [-.03, .06] | .01 [-.03, .06] | .04 [-.01, .09] |
| ***Mon. Enc.*** | .08*** [.05, .11] | .04* [.00, .07] | .09*** [.05, .12] | .04+ [-.00, .09] | .04+ [-.00, .09] | .08** [.02, .13] |
| ***Educ. * Verb. Enc.*** | | | | .05+ [-.01, .12] | -.02 [-.08, .04] | .04 [-.02, .11] |
| ***Educ. * Mon. Enc.*** | | | | .07* [.01, .13] | -.01 [-.08, .05] | .02 [-.05, .09] |
| ***$R^2$*** | 0.040 | 0.041 | 0.042 | 0.045 | 0.041 | 0.043 |
| ***Section B - Moderator: Civic knowledge*** | | | | | | |
| ***Civ. Know.*** | .15*** [.10, .20] | .15*** [.10, .20] | .18*** [.12, .23] | .05 [-.04, .14] | .07 [-.02, .16] | .16*** [.07, .26] |
| ***Verb. Enc.*** | .05** [.01, .08] | .00 [-.03, .03] | .06*** [.02, .09] | -.04 [-.13, .04] | -.06 [-.14, .03] | .01 [-.08, .10] |
| ***Mon. Enc.*** | .08*** [.05, .11] | .04* [.01, .07] | .09*** [.06, .13] | .01 [-.08, .09] | -.04 [-.13, .04] | .12** [.03, .21] |
| ***Civ. Know. * Verb. Enc.*** | | | | .15* [.03, .27] | .09 [-.04, .21] | .08 [-.06, .21] |
| ***Civ. Know. * Mon. Enc.*** | | | | .13* [.00, .25] | .14* [.01, .26] | -.05 [-.18, .09] |
| ***$R^2$*** | 0.052 | 0.041 | 0.064 | 0.058 | 0.045 | 0.068 |

Note: Section A reports models with educational attainment, and Section B with civic knowledge, across three policy topics. The first three columns present direct effect models (main effects only), and the last three display interaction models testing for moderation effects. Brackets indicate 95% CI. Asterisks denote significance (+p<.1, *p < .05, **p < .01, ***p < .001). DV ranges 0 (all incorrect) to 1 (all correct). See Appendix E for models with controls.

Restricting to tracked participants, we estimate the local average treatment effect (LATE, Table 2). Using instrumental variable regressions, we find a statistically significant effect of online information search on knowledge for child support and cannabis legalization (H2a). Consistent with ITT models, we find support for the knowledge gap hypothesis (H2b) for child support: individuals with higher education levels (educational attainment and civic knowledge) gained more knowledge compared to those with lower education levels.

**Table 2. Local average treatment effects (LATE) with and without interaction terms.**

| *Section A - Moderator: Educational attainment* | | | | | | |
|---|---|---|---|---|---|---|
| | **Child Support (N=702)** | **Energy Transition (N=637)** | **Cannabis Legalization (N=620)** | **Child Support (N=702)** | **Energy Transition (N=637)** | **Cannabis Legalization (N=620)** |
| ***Info-search*** | .24*** [.13, .35] | .12 [-.04, .27] | .43*** [.19, .68] | .08 [-.08, .24] | .11 [-.09, .31] | .31+ [-.01, .63] |
| ***Edu*** | .05** [.02, .09] | .06*** [.03, .10] | .06** [.02, .10] | -.02 [-.09, .04] | .06+ [-.01, .13] | .01 [-.09, .12] |
| ***Info-search * Edu*** | | | | .32** [.09, .55] | .01 [-.30, .31] | .25 [-.24, .75] |
| ***First-Stage F-Stat (Main)*** | 48.82 (p<.001) | 23.47 (p<.001) | 15.21 (p<.001) | 24.51 (p<.001) | 11.99 (p<.001) | 7.73 (p<.001) |
| ***Wu-Hausman*** | 6.24 (p=0.013) | 2.51 (p=0.114) | 10.48 (p=0.001) | 5.07 (p=0.007) | 1.26 (p=0.285) | 5.34 (p=0.005) |
| ***Sargan*** | 0.03 (p=0.873) | 0.54 (p=0.463) | 0.37 (p=0.545) | 0.18 (p=0.912) | 0.61 (p=0.738) | 1.07 (p=0.586) |
| ***First-Stage F-Stat (Info-search x Edu)*** | | | | 20.64 (p<.001) | 9.98 (p<.001) | 5.84 (p<.001) |
| *Section B - Moderator: Civic knowledge* | | | | | | |
| ***Info-search*** | .25*** [.14, .36] | .13+ [-.02, .28] | .46*** [.21, .70] | -.05 [-.33, .23] | -.08 [-.48, .32] | .36 [-.25, .97] |
| ***Civ. Know.*** | .13*** [.06, .19] | .12*** [.05, .19] | .13** [.05, .22] | -.01 [-.15, .13] | .06 [-.08, .20] | .12 [-.10, .33] |
| ***Info-search * Civ. Know.*** | | | | .50* [.05, .95] | .32 [-.26, .90] | .10 [-.95, 1.15] |
| ***First-Stage F-Stat (Main)*** | 50.05 (p<.001) | 24.71 (p<.001) | 15.62 (p<.001) | 25.23 (p<.001) | 12.57 (p<.001) | 8.30 (p<.001) |
| ***Wu-Hausman*** | 7.87 (p=0.005) | 3.75 (p=0.053) | 12.44 (p<.001) | 4.89 (p=0.008) | 2.60 (p=0.075) | 5.03 (p=0.007) |
| ***Sargan*** | 0.01 (p=0.911) | 0.66 (p=0.416) | 0.26 (p=0.609) | 0.49 (p=0.784) | 1.09 (p=0.580) | 5.01 (p=0.082) |
| ***First-Stage F-Stat (Info-search x Civ. Know.)*** | | | | 20.66 (p<.001) | 11.68 (p<.001) | 5.85 (p<.001) |

Note: The table reports instrumental variable (IV) regressions using random assignment to the information-search treatment as an instrument and its interaction with the moderator as excluded instruments.Sections A and B present models with educational attainment and civic knowledge as moderators. Instrument strength is assessed using first-stage F-statistics, reported for both the main endogenous regressor and the interaction term. The first three columns present direct effect models (main effects only), while the last three display interaction models testing for moderation effects. Brackets indicate 95% CI. Asterisks denote significance (+p < .10, *p < .05, **p < .01, ***p < .001). DV ranges 0 (all incorrect) to 1 (all correct). See Appendix F for models with controls.

To understand the mechanisms of the intervention, we explored whether the encouragement to inform affected information search differently (RQ3). We find statistically significant interactions of both conditions with civic knowledge across all policy topics, but not with educational attainment. All the significant coefficients are negative, i.e., the treatment effect is higher for individuals with lower civic knowledge; for educational attainment, most coefficients are also negative. Table 3 presents a summary.

**Table 3. Logistic regression of information search on education and intervention.**

| *A - Moderator: Educational attainment* | | | | | | |
|---|---|---|---|---|---|---|
| | **Child Support (N=702)** | **Energy Transition (N=637)** | **Cannabis Legalization (N=620)** | **Child Support (N=702)** | **Energy Transition (N=637)** | **Cannabis Legalization (N=620)** |
| ***Educ.*** | .23<br>[-.14, .60] | .10<br>[-.30, .49] | .15<br>[-.26, .56] | 1.36*<br>[.18, 2.86] | .24<br>[-.75, 1.29] | .78<br>[-.25, 1.96] |
| ***Verb. Enc.*** | 1.86***<br>[1.28, 2.49] | 1.15***<br>[.58, 1.76] | 1.15***<br>[.58, 1.77] | 2.73***<br>[1.66, 4.19] | 1.12*<br>[.26, 2.09] | 1.63**<br>[.69, 2.76] |
| ***Mon. Enc.*** | 2.44***<br>[1.88, 3.06] | 1.81***<br>[1.25, 2.41] | 1.53***<br>[.96, 2.14] | 3.27***<br>[2.21, 4.71] | 2.01***<br>[1.18, 2.96] | 1.98***<br>[1.05, 3.10] |
| ***Educ. * Verb. Enc.*** | | | | -1.32+<br>[-2.91, .01] | .06<br>[-1.16, 1.25] | -.78<br>[-2.12, .44] |
| ***Educ. * Mon. Enc.*** | | | | -1.22+<br>[-2.80, .07] | -.37<br>[-1.56, .78] | -.73<br>[-2.05, .47] |
| ***McFadden $R^2$*** | 0.126 | 0.071 | 0.053 | 0.131 | 0.073 | 0.056 |
| *B - Moderator: Civic knowledge* | | | | | | |
| ***Civ. Know.*** | .82*<br>[.10, 1.55] | 1.20**<br>[.39, 2.06] | .56<br>[-.27, 1.42] | 6.59***<br>[3.05, 10.89] | 4.76**<br>[1.87, 8.31] | 3.83*<br>[1.16, 7.04] |
| ***Verb. Enc.*** | 1.88***<br>[1.31, 2.52] | 1.18***<br>[.61, 1.80] | 1.16***<br>[.59, 1.78] | 6.44***<br>[3.42, 10.27] | 4.30**<br>[1.79, 7.44] | 3.93**<br>[1.65, 6.73] |
| ***Mon. Enc.*** | 2.47***<br>[1.91, 3.09] | 1.86***<br>[1.31, 2.47] | 1.55***<br>[.98, 2.16] | 7.27***<br>[4.28, 11.08] | 4.74***<br>[2.27, 7.85] | 4.17**<br>[1.93, 6.95] |
| ***Civ. Know. * Verb. Enc.*** | | | | -5.93**<br>[-10.37, -2.18] | -4.17*<br>[-7.92, -.96] | -3.84*<br>[-7.29, -.82] |
| ***Civ. Know. * Mon. Enc.*** | | | | -6.33**<br>[-10.73, -2.63] | -3.80*<br>[-7.51, -.65] | -3.62*<br>[-7.03, -.66] |
| ***McFadden $R^2$*** | 0.13 | 0.084 | 0.055 | 0.147 | 0.095 | 0.066 |

Note: The table reports interaction models testing whether the effects of verbal and monetary encouragement on information search vary by educational attainment (Section A) and by civic knowledge (Section B). They correspond to the IV first-stage regressions underlying Table 2, but are estimated here using logistic regression models to aid interpretation of the binary outcome. The first three columns present direct effect models (main effects only), while the last three columns display interaction models testing for moderation effects. Brackets indicate 95% confidence intervals. Asterisks denote significance (+p < .10, *p < .05, **p < .01, ***p < .001). See Appendix G for models with controls.

Figure 3 illustrates our results so far. The top row shows the effect of online search on knowledge scores (IVR; Table 2): in general, it appears that among participants who searched, those with higher educational attainment and civic knowledge achieved larger gains in knowledge. For those with low levels of educational attainment, there were no significant differences between individuals who searched for information and those who did not, across all policy topics; however, for cannabis legalization, treated participants outperformed controls in the ITT model ($EMM_{diff}$ = 0.06, CI [0.01, 0.10], p = .012), particularly in the monetary group ($EMM_{diff}$ = 0.08, CI [0.01, 0.14], p = .013). In contrast, participants with low levels of civic knowledge who searched for information did score significantly higher than those who did not, in the child support wave ($EMM_{diff}$ = 0.18, CI [0.05, 0.30], p = .005) and in the cannabis legalization wave ($EMM_{diff}$ = 0.40, CI [0.16, 0.65], p = .001); but, as with educational attaiment, significant differences were only found for the monetary group (child support: $EMM_{diff}$ = 0.11, CI [0.06, 0.16], p < .001; cannabis legalization: $EMM_{diff}$ = 0.08, CI [0.03, 0.13], p < .001). Appendix H reports full contrast tables for ITT and IVR.

**Figure 3. Interaction Effects on Knowledge Scores and Information Search.**

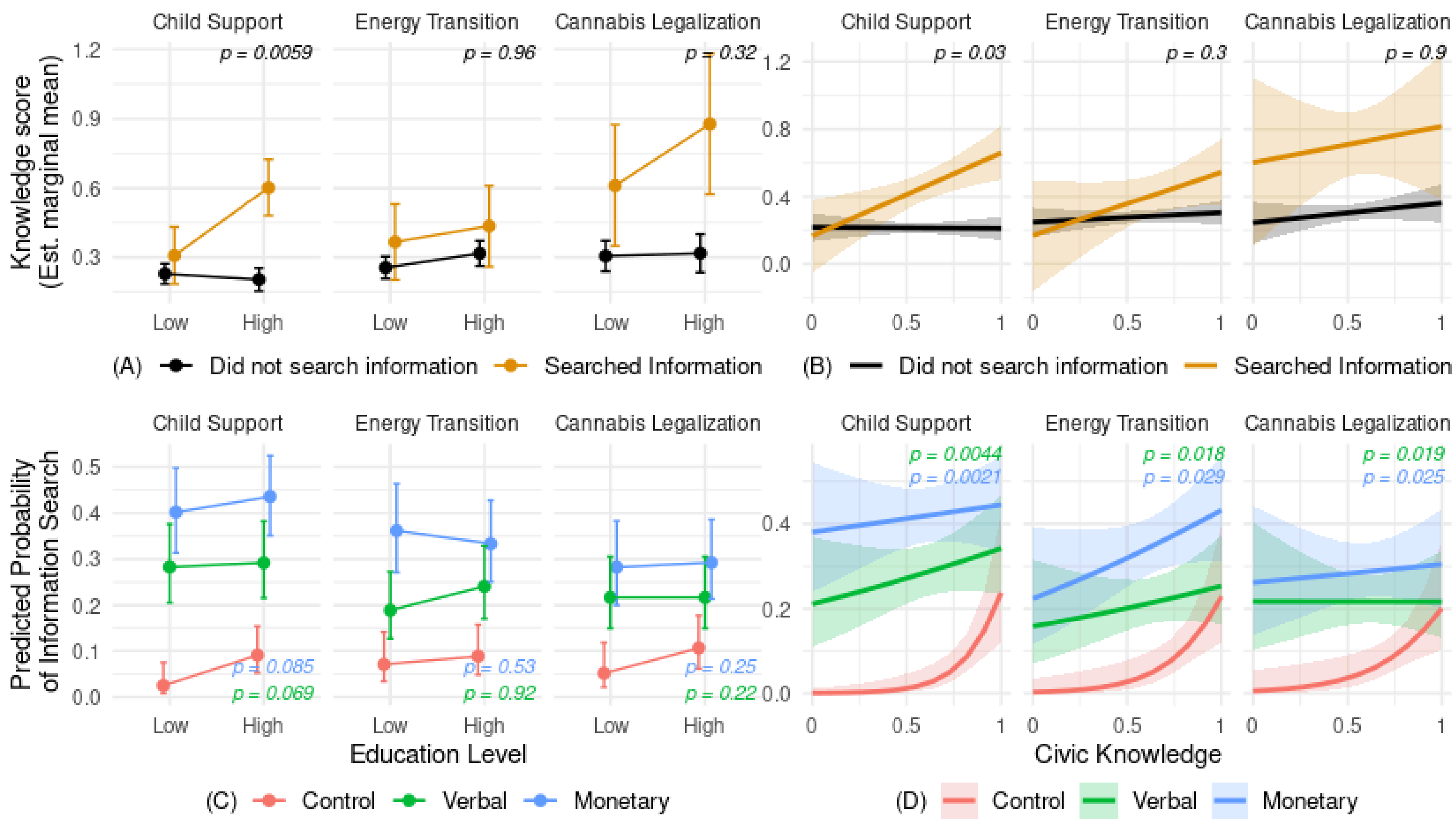

Note: The top row displays estimated marginal means of the knowledge score (y-axis) as a function of formal educational attainment (A) and civic knowledge (B), separately for participants who did or did not conduct an online information search (instrumented; see Table 2). The bottom row depicts the predicted probability of initiating an information search (y-axis) by education (C) and civic knowledge (D), disaggregated by experimental condition (control, verbal encouragement, monetary encouragement), based on binomial regressions (Table 3). Shaded areas represent 95% confidence intervals; p-values for interaction terms are shown in each panel.

[**Alt Text:** Multi-panel results for three policy topics: the top row plots knowledge scores by education and civic knowledge for searchers versus non-searchers, and the bottom row plots search probability by education and civic knowledge across experimental conditions.]

The bottom row of Figure 3 depicts the determinants of search behavior itself (see Table 3). In the encouragement conditions, there are no discernible differences in search likelihood by either educational attainment or civic knowledge; hence, the observed interaction effect stems entirely from the control (baseline) group, where individuals with higher educational attainment and more politically knowledgeable individuals are slightly more likely to initiate a search without being prompted to do so.

Beyond searching (or not) for information related to the policy, we describe the effort of information-seeking behaviors (RQ4, Table 4). We observe that while individuals with higher educational levels on average visited more pages (M = 2.66 vs. 1.86 pages; $p = .002$) and displayed more search events (M = 2.40 vs. 2.04 pages; $p = .090$), they seemingly spent less time on search pages (M = 31.09 s vs. 43.93 s; $p = .079$). The directionality of the correlations with civic knowledge echoes these patterns, though most correlations are not statistically significant: page visits (r = .09, $p$=0.099) and searches (r = .05, n.s.) correlate positively, while time per search correlates negatively (r = -.07, n.s.).

**Table 4. Information search effort by education level and civic knowledge.**

| | **Education** | | | **Civic Knowledge** | |
|---|---|---|---|---|---|
| | **Low** | **High** | **p** | **pearson r** | **p** |
| ***Policy-related URL page visits*** | 1.86 [1.64; 2.09] | 2.66 [2.21; 3.11] | **0.002** | 0.09 | *0.099* |
| ***Median time per policy-related URL page visit*** | 127.34 [102.21; 152.46] | 128.43 [105.20; 151.65] | 0.950 | -0.02 | 0.771 |
| ***Policy-related searches*** | 2.04 [1.75; 2.34] | 2.40 [2.11; 2.68] | *0.090* | 0.05 | 0.352 |
| ***Median time per policy-related search*** | 43.93 [30.72; 57.14] | 31.09 [25.48; 36.70] | *0.079* | -0.07 | 0.171 |

Note: The first column lists the observed metrics of information-seeking effort. Columns two to four compare metrics between individuals with low and high educational attainment using paired t-tests. Columns five and six report Pearson correlation coefficients between each metric and civic knowledge. See Appendix I for tables per wave.

## Discussion

Building on the knowledge gap literature, this study investigates the effect of online information search, a prototypical case of intentional behaviour, on political knowledge. First, we conceptualize mechanisms associated with knowledge inequalities, including those that emerge due to digital environments (Figure 1). Informed by this framework, we analyzed the results of an experiment in which participants were encouraged to search for online information on three salient policy topics, diverging in divisiveness and complexity (center block). While our encouragement directly targets motivation to search information, it does not alter the structural and other individual mechanisms (left and right blocks). Beyond educational attainment, the canonical moderator in knowledge-gap research, we also examine civic knowledge as a domain-specific capacity for political learning. In addition to the confirmatory assessment of the knowledge gap hypothesis, we analyze the direct effects of the intervention on search behaviour and an exploration of the associations between education and online search effort.

We find support for the knowledge gap hypothesis. For basic child support, we find an increase in the knowledge disparity between individuals with lower and higher education (measured by either educational attainment or civic knowledge) due to information search. The intention-to-treat (ITT) estimates indicate that the interaction between the intervention encouragements and educational levels accounts for an increase in the knowledge gap of 7 to 15 percentage points (Table 1). While these immediate effect sizes might appear modest in a single snapshot, they are substantively critical because knowledge acquisition is cumulative; participants with higher baseline (civic) knowledge gained more from the search task, supporting the idea that disparities widen over repeated daily media interactions. Furthermore, the ITT estimates are conservative because they include individuals who did not engage in the search task despite the encouragement (non-compliers).

When isolating the effect for those who actually conducted a search (LATE), the effect is more pronounced, ranging between 32 and 50 percentage points for the child support topic (Table 2), representing a gain of approximately one to two additional correct answers on the five-item test after a single search session. This finding is noteworthy because our behavioral data confirms that the intervention effectively leveled the ground in terms of information-seeking engagement. As illustrated in Figure 3, individuals with higher education were more likely to search without being prompted to do so (in the control group), but the encouragements overrode these initial differences in intrinsic motivation, resulting in similar search rates across educational levels. Despite equalized information exposure, the knowledge differentials persisted or, in some cases, widened.

In the wave focused on the energy transition policy, we also found individuals with more civic knowledge to gain more knowledge when encouraged to search in the monetary condition (compared to those with less civic knowledge); however, the corresponding interaction of civic knowledge with information search (i.e., LATE) was non-significant. We attribute this divergence to the complexity of this specific topic, where a solidified baseline of political knowledge is likely more relevant for successful learning than the gains achievable in a one-shot search task. We did not find evidence for the knowledge gap hypothesis in the wave related to cannabis legalization.

Across all topics and independent variables, we don't find statistical support for the knowledge gap's counter-hypothesis, i.e., that participants with lower educational attainment or civic knowledge close the gap under our directives to search: the knowledge differential either persists or, as seen before, increases after the intervention. While previous research has shown leveling effects when experimentally introducing incidental exposure (e.g., Leeper, 2020), our findings underscore that learning processes differ under active, self-directed search conditions.

Our results complement Leeper's conclusion that knowledge differentials arise from who gets exposed (compared to who learns better); it is plausible that the exposure to self-selected content can also play a crucial role when searching in high-choice environments.

Although we expected knowledge gains across all educational levels, knowledge gains among participants with low education or low civic knowledge were limited. All significant improvements for these groups emerged only in the monetary condition, suggesting that intrinsic motivation (activated by a verbal encouragement) alone might be insufficient, and that external constraints may hinder engagement unless a tangible external reward is offered. However, it is also in the monetary condition that the knowledge gap hypothesis is most supported, suggesting that, among higher-education participants, the benefits of the verbal and monetary conditions are accumulated. However, while the use of monetary incentives is theoretically supported (Prior & Lupia, 2008), we did not find statistical differences in the direct contrast between the verbal and monetary encouragement. Among participants with low educational attainment, we find knowledge gains only for the policy that had been rated as least complex (cannabis legalization). Those with low civic knowledge, by contrast, also scored higher when tested on child support, a topic of moderate complexity and high personal salience. We didn't hypothesize different expectations for educational attainment and civic knowledge, but it is possible that the divergence in results might be due to associations with distinct mechanisms: formal education tracks structural resources (general skills and motivation), while civic knowledge reflects applied political capacities. We found no evidence that they were more familiar with the specific topics, see top row of Figure 3. Hence, while incentives may benefit some groups, skill-building is likely necessary for structurally disadvantaged groups who may lack the competencies to select relevant sources from among the ranked options presented by platforms. Moreover, algorithms may inadvertently foreground sources that are ill-suited for some individuals; for instance, in their attempts to prioritize quality[2], platform designers may overlook complexity.

Regarding the characteristics of the topics, our results align with findings by Lind and Boomgaarden (2019), who show that the knowledge gap is more pronounced for uncontested topics, i.e., policies that are not the subject of polarized public debate. In contested topics, value alignment may override information-based learning, either by motivating the learning process or because individuals are more familiar with the topic from the outset; a mechanism consistent with identity-driven patterns of belief formation (Hindman, 2009). The significant interaction between civic knowledge and the monetary encouragement for the energy transition indicates that, in line with Kleinnijenhuis (1991), topic complexity is also a moderator of the knowledge gap; though the effect was exclusive to the ITT for our intervention. Our findings thus suggest that while complexity remains a significant barrier to learning, the uncontested (non-polarizing) nature of a topic may be a more relevant precondition for the manifestation of knowledge gaps during self-directed search.

Exploratively, we found descriptive differences in metrics of the effort of information search; individuals with lower educational levels visited fewer pages and might (non-significant trend) have taken longer to select their source if they used a search engine. To be sure, sample sizes here restrict the analysis as only those who searched can be considered. Nevertheless, the findings align with previous research (Leeper, 2020) that media habits (and also interest and motivation), rather than heterogeneity effects from learning capabilities, were the cause of the found knowledge gaps. Future studies should investigate whether behavioral patterns differ between educational levels. Should the trend be replicated, it might indicate that individuals with higher levels of education possess abilities that allowed them to be more efficient in their information-seeking approach (see online search skills, Figure 1), maybe because they have adapted better to the structural changes in the media environment and are more practiced in similar tasks.

[2] https://www.google.com/intl/en_us/search/howsearchworks/how-search-works/ranking-results/

While the knowledge gap is often presented as a broad phenomenon that unfolds over extended periods, this study provides an analysis of mechanisms by examining the role of information seeking and improves on the validity by using behavioral data, circumventing the artificial constraints of controlled media environments common in other research designs. Unlike most experimental studies that limit information effects to specific media sources, such as curated articles or predefined media channels, we allowed participants to self-direct their searches and decide on the sources, with no control or curation by the researchers. Instead, participants were confronted with the restrictions imposed by the contemporary online environment, e.g., the options provided by the search engine of their preference. In doing so, we directly link self-selected source choices to political knowledge scores in an experimental trial. Our results underscore an inherent tension in new media contexts: free and extensive information, while democratically appealing, may disproportionately benefit those already equipped with stronger educational and cognitive resources. Thus, addressing knowledge gaps effectively may require interventions that go beyond simply increasing information accessibility or even direct encouragements, emphasizing the need to design digital infrastructures tailored to enhance the engagement and competencies of individuals with fewer initial advantages and extensive media habits.

### Limitations

Our sample reflects a highly specific population: members of a commercial web-tracking panel based in Germany. While the setting enables more precise behavioral measurement, it imposes clear limits on generalizability. Panel members are, on average, older, more educated, and more likely to be male than the broader German population, and they opted into both panel participation and web tracking, which may indicate higher levels of digital literacy than the general public. These factors raise the possibility that observed treatment effects may differ in younger, less educated, or less digitally engaged populations, or in countries with different political and media systems. Additionally, although all participants had opted into tracking months prior to the experiment, and many had long-standing panel membership, we cannot fully rule out the possibility that awareness of being observed subtly shaped behavior during the study window.  Therefore, the study does not claim that findings are representative of the broader German population, nor should results be generalized beyond the specific timeframe or policy context of examined here, but should be taken as a starting and comparison point for further investigation.

The behavioral measures covered both desktop and mobile browser-based activity, allowing for granular observations of web use across devices, which remains relatively uncommon in digital trace studies, but it only reflects exposure through web-based information environments as it can miss participants who sought information via other devices, native mobile applications, or offline sources. This limitation would specifically affect the exclusion restriction (Appendix C) and, thus, LATE estimates (i.e., Table 2): the instrumental-variable approach may assign some policy learning to observed online search that occurred via other pathways. Therefore, LATE estimates should be interpreted as capturing encouragement-induced information acquisition as reflected in tracked search behavior.

## Conclusion

This study distinguishes individual and structural features of the information environment and shows that explicit encouragement to search online in a naturalistic setting does not eliminate existing knowledge gaps, as advantaged participants still achieved higher knowledge gains. Reducing these disparities will require interventions that both strengthen individual capacity to process political information and adapt digital environments to better support learning among less advantaged groups.

## Ethics statement

The study received ethical approval from the Institutional Review Board (IRB) of the University of Konstanz (protocol IRB23KN02-003/w). Informed consent was obtained from all participants at every point of contact.

**Data availa**bility statement

Data and analysis code are publicly available via the Open Science Framework at https://osf.io/pv8ey/files/osfstorage?view_only=f86afcdc082b4814beed8130a9f33d5c. The study was preregistered on AsPredicted (https://aspredicted.org/td36w.pdf; see Appendix K). The author responsible for preregistration did not have access to the data prior to submission of the preregistration.